# Recoil effects in microwave Ramsey spectroscopy

Peter Wolf [1]
*BNM-SYRTE, Observatoire de Paris, 61 av. de l'Observatoire, 75014 Paris, France*
*Bureau International des Poids et Mesures, Pavillon de Breteuil, 92312 Sèvres, Cedex, France*

Christian J. Bordé
*Laboratoire de Physique des Lasers, Université Paris-Nord, 99 av. J.-B. Clément, 93430 Villetaneuse, France*

**Abstract:** *We present a theory of recoil effects in two zone Ramsey spectroscopy, particularly adapted to microwave frequency standards using laser cooled atoms. We describe the atoms by a statistical distribution of Gaussian wave packets which enables us to derive and quantify effects that are related to the coherence properties of the atomic source and that have not been considered previously. We show that, depending on the experimental conditions, the expected recoil frequency shift can be partially cancelled by these effects which can be significant at microwave wavelengths whilst negligible at optical ones. We derive analytical expressions for the observed interference signal in the weak field approximation, and numerical results for realistic caesium fountain parameters. In the near future Cs and Rb fountain clocks are expected to reach uncertainties which are of the same order of magnitude ($10^{-16}$) as first estimates of the recoil shift at microwave frequencies. We show, however, that the partial cancellation predicted by the complete theory presented here leads to frequency shifts which are up to an order of magnitude smaller. Nonetheless observation of the microwave recoil shift should be possible under particular experimental conditions (increased microwave power, variation of atomic temperature and launching height etc.). We hope that the present paper can provide some guidance for such experiments that would test the underlying theory and its assumptions, which in turn is essential for the next generation of microwave frequency standards.*

## INTRODUCTION

Generally speaking, any superposition of internal energy states of an atom induced by interactions with an electromagnetic field will give rise to a superposition of external states whose momenta will differ by $\hbar \mathbf{k}$ (where $\hbar$ is Planck's constant and $\mathbf{k}$ is the wave vector of the electromagnetic field) due to the absorption of the photon momentum by the atom. The different external states will lead to additional energies (kinetic, potential), which will appear as a frequency shift of the atomic transition frequencies $\omega$ measured in atomic spectroscopy. The additional kinetic energy gives rise to the well known recoil shift $\delta$

$$\frac{\delta}{\omega} = \frac{(\hbar k)^2}{2M} \frac{1}{\hbar \omega} = \frac{\hbar \omega}{2Mc^2} \tag{1}$$

for the case of a freely propagating wave ($k = \omega/c$) with $M$ the mass of the atom and $c$ the velocity of light in vacuum. The relative magnitude of the recoil shift is proportional to the transition frequency so that its observation becomes more difficult as the wavelength of the transition increases, which has rendered its observation in the microwave domain impossible so far in spite of the excellent performance of microwave clocks and frequency standards (as an example, (1) leads to a shift of only $\approx 1.5 \times 10^{-16}$ for the hyperfine transition of caesium at 9.2 GHz). On the other hand the recoil shift is easily observed for X and $\gamma$ rays where it is much larger than the natural line widths, and has first been observed in Doppler-free optical spectroscopy 30 years ago [1].

However, microwave frequency standards have drastically improved in recent years due to the advent of cold atom fountain clocks which now routinely reach uncertainties of one part in $10^{15}$ and slightly below [2 - 5]. In the near future new methods for controlling the major systematic effects [6] and the advent of space based clocks like the ACES mission (planned for 2008 [7]) should push the uncertainties to the low $10^{-16}$ regime or below. Then the $10^{-16}$ order of magnitude of the recoil shift is well within the sensitivities of such devices and detailed predictions of its magnitude and its variation with experimental parameters will be required.

It turns out that the detailed theory of the recoil shift for cold atoms in the microwave domain is richer than its optical counterpart. This is due to the fact that the interactions take place in standing waves (in microwave cavities) with the microwave wavelength comparable to the characteristic size of the atomic cloud. On one hand the standing waves lead to multiple photon processes (see e.g. [8]), with atoms absorbing photons

[1] part of this work was carried out under CNES research grants 793/00/CNES/8201 and 793/02/CNES/480000078

from one travelling wave component of the field and emitting them into another. This leads to final external states with total momenta $n\hbar k$ (where n is an integer) so the order of magnitude of the shift obtained from (1) may only be a lower limit of the total shift. On the other hand, the comparable size of the electro-magnetic wavelength and of the atomic cloud give rise to non negligible contributions with opposite recoil shifts (cf. section 3), as discovered only recently [9, 10], which may lead to significant cancellations. So the observed total frequency shift is a result of the competition of these two effects that depends critically on the particular experimental parameters, in particular on the field strengths (microwave power) and the characteristic parameters of the external states of the atoms.

In this paper we provide a first complete theory of recoil effects in microwave Ramsey spectroscopy, and more generally, in two-zone interferometers using standing waves. To do so we use a description of the atoms as a statistical ensemble of Gaussian wave packets which allows us to take into account both of the above effects in realistic experimental situations. The aim is to provide some guidance for experiments that could measure the effects and thereby confirm or contradict the underlying theory and its assumptions, which in turn is essential for the next generation of microwave frequency standards.

The pioneering theoretical work for the description of recoil effects in saturated optical spectroscopy is due to [11], with several other approaches and refinements published subsequently [1, 12, 13]. The strong field case in optical standing waves, taking into account multiple photon processes, was considered extensively in [8]. However, that paper did not consider components with opposite recoil velocities which are responsible for partial cancellation of the recoil shift (c.f. section 3), which is justified in the optical case where they play no significant role. Such effects were discovered only recently [9, 10] when using the methods of Gaussian optics in atom inteferometry for the first studies of the microwave recoil. Indeed, recently the experimental and theoretical developments in the domains of atomic spectroscopy and atomic inertial sensors have gone hand in hand as they present two facets of the same fundamental processes (see e.g. [15] for a comprehensive review].

Much of the early theory of clocks and atom interferometers describes the external states of the atoms either classically [16] or as plane waves [17, 18]. More recently, the use of Gaussian wave packets for the external states [15, 19, 20] and its application to microwave frequency standards [9, 10] has led to significant new results, in particular, the predictions of a partial cancellation of the recoil shift. In this paper we use such an approach but extend it to a statistical ensemble of such wave packets (describing the atomic cloud) which in turn leads to predictions in terms of known parameters (temperature and size of the atomic cloud) that can differ significantly from those of [9, 10].

Similar to the approach taken in [9, 10] we describe the external state of the individual wave packets as Gaussian in the two spatial dimensions that are perpendicular to the atomic trajectory but classically in the dimension parallel to the atomic trajectory. This provides a description of the wave-packets in terms of characteristic times (end of cooling, passage in the cavities, detection) rather than characteristic positions along the atomic trajectory. A more rigorous approach which does not require such a classical approximation can be found in [8]. For the purposes of the present paper the two approaches lead to similar results and predictions for the observable effects in "standard", present day microwave frequency standards (section 4). A discussion of the relation between the two approaches in the context of recoil effects in microwave frequency standards will be the subject of a forthcoming publication.

We introduce our description of the atoms in section 1 using standard Gaussian wave packets and a density operator that describes the statistical ensemble of such wave packets, characterised by the statistical distributions of their central velocities and initial positions. We link these distributions and the characteristic width of the individual wave packets to the measured parameters of the atoms (temperature and initial size of the cloud). In section 2 we provide the general theory used to describe the interactions with the microwave fields, the free propagation, and the detection of the atoms. In the weak field approximation (section 3) we find analytical solutions to the differential equations and the integrals of the general theory of the previous section, resulting in analytical expressions for the final detection probability (and hence the recoil frequency shift) in terms of the experimental parameters. Finally we study the realistic strong field case by numerical simulation in section 4. We give results for "typical" microwave fountain parameters showing the calculated frequency shift and fringe contrast as a function of the experimental parameters (temperature, launching height, microwave power, initial width of the cloud, etc.). We provide a qualitative understanding of these results by comparing them to the analytical results of the previous section. We conclude with a summary and a discussion of related issues that could be the subject of future work in this field.

## 1. ATOMIC WAVE FUNCTION AND DENSITY MATRIX

A treatment of atom interferometers (including frequency standards) that takes into account the external degrees of freedom of the atoms requires a quantum description of the atom's internal and external states [14]. Recently, methods based on the description of the atoms as Gaussian wave packets and their propagation using the ABCD formalism of Gaussian optics were developed [15, 20]. In this work we use such a Gaussian description, i.e. we approximate the atomic cloud as a statistical ensemble of Gaussian wave packets of identical waist (minimum spatial width) with a statistical distribution of their central velocities and positions. It turns out that for the present work such a description is indispensable as it leads to results that are significantly different from the ones obtained using plane waves or a single Gaussian wave packet (see sections 3 and 4).

In practice, the atoms are laser cooled in a magneto optical trap (MOT) or an optical molasses and then released (launched) to propagate freely through the microwave interaction regions. The temperature (velocity distribution) of the atoms is determined by the time of flight technique and is typically less than a few µK. The initial width of the atomic cloud (measured initial position distribution) is determined by imaging and typically ranges from 1 mm (MOT) to 5 mm (molasses).

We describe the external state of each individual atom in terms of a Gaussian wave packet, a well known solution of the free particle Schrödinger equation for a particle of mass $M$ [21, 22]. The wave function of such a wave packet centred around a position $z_i$ at $t = 0$ and a velocity $v_i$ can be written

$$\varphi(t,z) = N(t)\exp\left[(-P(t) + iQ(t))(z - (z_i + v_i t))^2\right]\exp\left[\frac{iM}{\hbar}\left(v_i(z - z_i) - \frac{v_i^2}{2}t\right)\right] \tag{2}$$

where the normalisation function is

$$N(t) = \frac{e^{-i\vartheta(t)}}{\left[2\pi(\Delta z)^2\left(1 + \frac{1}{\chi}\right)\right]^{1/4}} \tag{3}$$

$$\text{tg}(2\vartheta) = \frac{\hbar t}{2M(\Delta z)^2} \tag{4}$$

and the functions describing the spreading of the wave packet and its phase curvature are

$$P(t) = \frac{M^2(\Delta z)^2}{\hbar^2 t^2(1+\chi)} \tag{5}$$

$$Q(t) = \frac{M}{2\hbar t(1+\chi)} \tag{6}$$

with $\chi$ defined as

$$\chi = \frac{4M^2(\Delta z)^4}{\hbar^2 t^2} \tag{7}$$

and with $\Delta z$ the waist (spatial width at $t = 0$) of the wave packet. For simplicity we consider here only one dimension, the generalisation to a 3D Gaussian wave packet being straightforward.

The wave packet starts expanding when the trapping and cooling fields are switched off, we therefore take that moment to be the initial time $t = 0$. In practice the cooling fields are turned off slowly (adiabatic cooling phase) over a few 100 µs, leading to a distribution of initial times among the atoms which is negligible for the purposes of this paper (see section 3).

We describe the statistical ensemble of atoms with external states given by (2) using a density operator of the form

$$\rho(t) = Z_{p_i}^{-1} Z_{z_i}^{-1} \int dp_i \int dz_i q(p_i) s(z_i) |\varphi_{p_i z_i}(t)\rangle\langle\varphi_{p_i z_i}(t)| \tag{8}$$

where $|\varphi_{p_i z_i}(t)\rangle$ is a Gaussian wave packet with central momentum $p_i$ and initial central position $z_i$, i.e.

$$\langle p|\varphi_{p_i z_i}(t)\rangle = \left(\frac{2(\Delta z)^2}{\pi\hbar^2}\right)^{1/4}\exp\left\{-\frac{(\Delta z)^2(p-p_i)^2}{\hbar^2}\right\}\exp\left\{-\frac{i}{\hbar}\left(pz_i + \frac{p^2 t}{2M}\right)\right\} \tag{9}$$

and $\langle z | \varphi_{p_i z_i}(t) \rangle = \varphi(t,z)$ of equation (2) with $v_i = p_i/M$.

Two measurements are available: the momentum distribution (expressed in terms of the measured temperature $\theta$)

$$P(p) = \langle p | \rho(t) | p \rangle = \frac{1}{\sqrt{2\pi M k_B \theta}} \exp\left\{-\frac{p^2}{2M k_B \theta}\right\} \tag{10}$$

and the initial position distribution

$$P(z, t=0) = \langle z | \rho(t=0) | z \rangle = \frac{1}{\sqrt{2\pi w^2}} \exp\left\{-\frac{z^2}{2w^2}\right\} \tag{11}$$

where $k_B$ is the Boltzmann constant and $w$ is the measured initial width of the atomic cloud. Inserting the density operator (8) into the left hand side of condition (10) we deduce that the probability density $q(p_i)$ describing the distribution of central momenta $p_i$ is

$$q(p_i) = \exp\left\{-\frac{p_i^2}{2M k_B \theta - \frac{\hbar^2}{2(\Delta z)^2}}\right\} \tag{12}$$

with the corresponding normalisation constant

$$Z_{p_i} = \sqrt{\pi \left(2M k_B \theta - \frac{\hbar^2}{2(\Delta z)^2}\right)}. \tag{13}$$

Similarly when inserting (8) into the left hand side of condition (11) we deduce

$$s(z_i) = \exp\left\{-\frac{z_i^2}{2(w^2 - \Delta z^2)}\right\} \tag{14}$$

with the corresponding normalisation constant

$$Z_{z_i} = \sqrt{2\pi(w^2 - \Delta z^2)}. \tag{15}$$

The resulting density operator ((8) with (12) to (15)) corresponds to a statistical mixture of Gaussian wave packets and is characterised by three parameters: the temperature $\theta$, the initial width of the cloud $w$, and the waist of the wave packets $\Delta z$. Two of them are known from measurements ($\theta$ and $w$) the third one ($\Delta z$) being unknown. However, the momentum and initial position distributions of (8) with (12)-(15) are only defined under the condition $\hbar^2/(4M k_B \theta) \leq (\Delta z)^2 \leq w^2$. This reflects the fact that the waist of the individual wave packets cannot be larger than the measured initial width of the cloud, or its momentum counterpart ($\hbar/(2\Delta z)$) larger than the measured momentum distribution of the cloud. The equality describes the pure state where all individual wave packets have the same central velocities and initial positions ($q(p_i)$ and $s(z_i)$ are non zero only for $p_i=z_i=0$).

## 2. GENERAL THEORY

Quite generally we will consider atom interferometers with two spatially separated interaction zones (Ramsey geometry, as used in most atomic clocks). We do not consider phase shifts due to gravitational or inertial fields in this work (for a general treatment of atom interferometry including those see [15, 19, 20]). We will first describe the pure case (a single Gaussian wave packet (2)) and integrate at the end over the statistical distributions ((12) and (14)) of central velocities and positions.

The atom propagates along the x direction crossing the interaction zones (cavities) of spatially limited electromagnetic fields. The spatial cut-offs of the fields imply that the wave vector of the field inside the interaction zones is distributed over the width of the Fourier transform of the spatial form of the field ($\delta k \approx 1/l$ where $l$ is the spatial width of the field, i.e. the size of the cavity) in all three dimensions. Additionally the field is switched on and off i.e. it is limited in time and therefore its frequency inside the cavity is also distributed ($\delta f \approx 1/T_s$ where $T_s$ is the time during which the field is on). However, only the spatio-temporal form of the field actually "seen" by the atom will play a role, more precisely, the form of the field in space-time regions where the atomic wave function $\Psi(t,\mathbf{r}) \approx 0$, plays no role (mathematically this is reflected by the fact that the coupling field appears in the Schrödinger equation (20) only in a product with $\Psi(t,\mathbf{r})$). As a consequence the temporal limitations of the field are irrelevant provided it is switched on and off when the atom is well outside the interaction zone, and similarly the spatial limitations in the z and y directions are irrelevant provided the size of $\Psi(t,\mathbf{r})$ is smaller than the extension of the field in those directions. On the other hand $\Psi(t,\mathbf{r})$ will always overlap with the limits of the field in the x direction (the atom "sees" the field limits when entering and exiting the interaction zones) so for the x-component of the photon wave vector $\delta k_x \approx 1/l_x$. The $\mathbf{k}$ vector actually absorbed by the atom is then determined from energy-momentum conservation by

$$\mathbf{k} \cdot \mathbf{v}_i + \frac{\hbar k^2}{2M} = \Delta \tag{16}$$

where $\mathbf{v}_i$ is the initial group velocity of the atomic wave packet and $\Delta$ the detuning of the field from resonance ($\Delta = \omega - (E_e-E_g)/\hbar$). To satisfy (16) for any value of $\Delta$ the part of the atomic wave packet in the excited internal state after the transition will have an additional momentum $p_x(\Delta)$ in the x-direction. That in turn leads to a phase difference between the excited and ground states at the second interaction of order $p_x L/\hbar$ (where L is the distance between the interaction zones) giving rise to the interference (Ramsey) fringes observed in the interferometer (clock). So in this interpretation of the interference fringes for two spatially separated interaction regions the interference arises from the recoil component parallel to the atom trajectory [15]. For $\mathbf{v}_i$ along the x direction one obtains $k_x$ from (16) leading to the phase difference

$$\Delta\phi = \left(\Delta - \frac{\hbar k^2}{2M}\right)\frac{L}{v_x} \tag{17}$$

which is equivalent to the usual expression under the classical approximation (neglecting the recoil term and setting $L/v_x = T$ for classical point particles).

Throughout this paper we describe the motion of the atoms in the x-direction classically but quantify the motion perpendicular to the x-axis retaining thereby most of the recoil phase shift (see below). The equations corresponding to a quantized motion in the x-direction can be found in [8]. They lead to results that, for the purposes of this paper, are equivalent to the ones obtained below. Here we describe the position of the atom along the x-direction by $v_x t$ where $v_x$ is the x component of the initial group velocity of the wave packet so the interference signal will be obtained in the standard manner directly from the energy difference of the internal states and the free flight time $T = L/v_x$. Doing that we neglect recoil terms from the x component of the absorbed wave vector. For typical caesium fountain parameters ($\Delta \approx 1$ Hz, $v_x \approx 2.5$ m/s, $T \approx 0.5$ s, $\omega_{eg}/2\pi \approx 9.2$ GHz) the vector $\mathbf{k}$ is constrained by (16) to be perpendicular to $\mathbf{v}_i$ to within 1° or less so the neglected terms lead to relative frequency shifts of $10^{-19}$ or less.

The complete wave function of the atom can be expressed as a two component matrix for the ground and excited internal states

$$\Psi(t,\mathbf{r}) = \begin{pmatrix} \Psi_e(t,\mathbf{r}) \\ \Psi_g(t,\mathbf{r}) \end{pmatrix}. \tag{18}$$

The coupling field for the magnetic dipole transition used in microwave frequency standards is a standing magnetic field inside the microwave cavity, described classically by

$$B\, f(\mathbf{r}) \cos(\omega t + \phi) \tag{19}$$

where $B$ is the amplitude of the magnetic field component parallel to the static magnetic field with $\omega$ its angular frequency and $\phi$ an arbitrary initial phase. The function $f(\mathbf{r})$ describes the spatial form of the standing wave. We assume a near resonant field and use the rotating wave approximation, neglecting off-resonant terms at frequencies $2\omega$, so the Schrödinger equation reduces to a pair of coupled equations

$$i\frac{\partial}{\partial t}\Psi_e(t,\mathbf{r}) = -\frac{\hbar}{2M}\nabla^2\Psi_e(t,\mathbf{r}) + \frac{\Omega}{2}e^{-i\Delta t}f(\mathbf{r})\Psi_g(t,\mathbf{r})$$
$$i\frac{\partial}{\partial t}\Psi_g(t,\mathbf{r}) = -\frac{\hbar}{2M}\nabla^2\Psi_g(t,\mathbf{r}) + \frac{\Omega}{2}e^{i\Delta t}f(\mathbf{r})\Psi_e(t,\mathbf{r})$$
(20)

for free particles in the presence of the coupling field. Here $M$ is the mass of the atom, $\Delta = \omega - (E_e-E_g)/\hbar$ is the detuning and $\Omega$ is the Rabi frequency. For microwave frequency standards $\Omega \approx \mu_B B/\hbar$ [23] where $\mu_B$ is the Bohr magneton. Note that (20) reduces to the simple free particle Schrödinger equation in the absence of the coupling field ($\Omega = 0$) and to standard Rabi equations for a two level system when the spatial dependence is neglected ($\Psi(t,\mathbf{r}) \to \Psi(t)$ and $f(\mathbf{r}) = 1$).

In a square microwave cavity, as used in many primary frequency standards, the form of the standing wave can be written

$$f(\mathbf{r}) = V(x)\cos[k(z + (z_i + v_i t))]$$
(21)

where $V(x)$ describes the dependence parallel to the atom trajectory ($V(x) = \cos(k_x x)$ in the cavity frame). For the z-dependence we have used a frame centred on the initial position $z_i$ and initial velocity $v_i$ of the incident atomic Gaussian wave packet (cf. (2)). There is no dependence on y so we will neglect the y-dimension and, as mentioned above, treat the x-dimension classically so $V(x)$ simply leads to a time dependence of $\Omega$ in (20) ($\Omega\to\Omega(t)$, the instantaneous Rabi frequency "seen" by the atoms). Under those conditions only the z-dependence remains and (20) reduces to one dimension.

To solve (20) with (21) we try the substitution

$$\Psi_e(t,z) = \sum_{a,n} e_n^a(t)\varphi_n^a(t,z)$$
$$\Psi_g(t,z) = \sum_{a,n} g_n^a(t)\varphi_n^a(t,z)$$
(22)

where a,n are integers and $\varphi_n^a(t,z)$ are normalised Gaussian wave packets (cf. (2)) centred on velocities $nv_r$ and positions $av_r T$ at $t=T_b$ (i.e. $z_i = av_r T - nv_r T_b$ and $v_i = nv_r$ in (2)), each multiplied by a complex coefficient $e_n^a(t)/g_n^a(t)$ that depends only on time ($v_r = \hbar k/M$ is the recoil velocity). We denote $T_b$ the time elapsed between the end of the cooling phase ($t=0$) and the first interaction, $T$ the time interval between the two interactions, and $T_d$ the time of detection (see figure 1). Substituting (22) into (20) with (21) we note that the resulting equations are satisfied if the $\varphi_{e/g}(t,z)$ are solutions of the free particle Schrödinger equation (which is the case for Gaussians) and if

$$i\sum_{a,n}\dot{e}_n^a(t)\varphi_n^a(z,t) = \frac{\Omega(t)}{4}e^{-i\Delta t}(e^{ik(z+z_i+v_i t)} + e^{-ik(z+z_i+v_i t)})\sum_{a,n}g_n^a(t)\varphi_n^a(z,t)$$
$$i\sum_{a,n}\dot{g}_n^a(t)\varphi_n^a(z,t) = \frac{\Omega(t)}{4}e^{i\Delta t}(e^{ik(z+z_i+v_i t)} + e^{-ik(z+z_i+v_i t)})\sum_{a,n}e_n^a(t)\varphi_n^a(z,t)$$
(23)

where a dot denotes differentiation with respect to $t$.

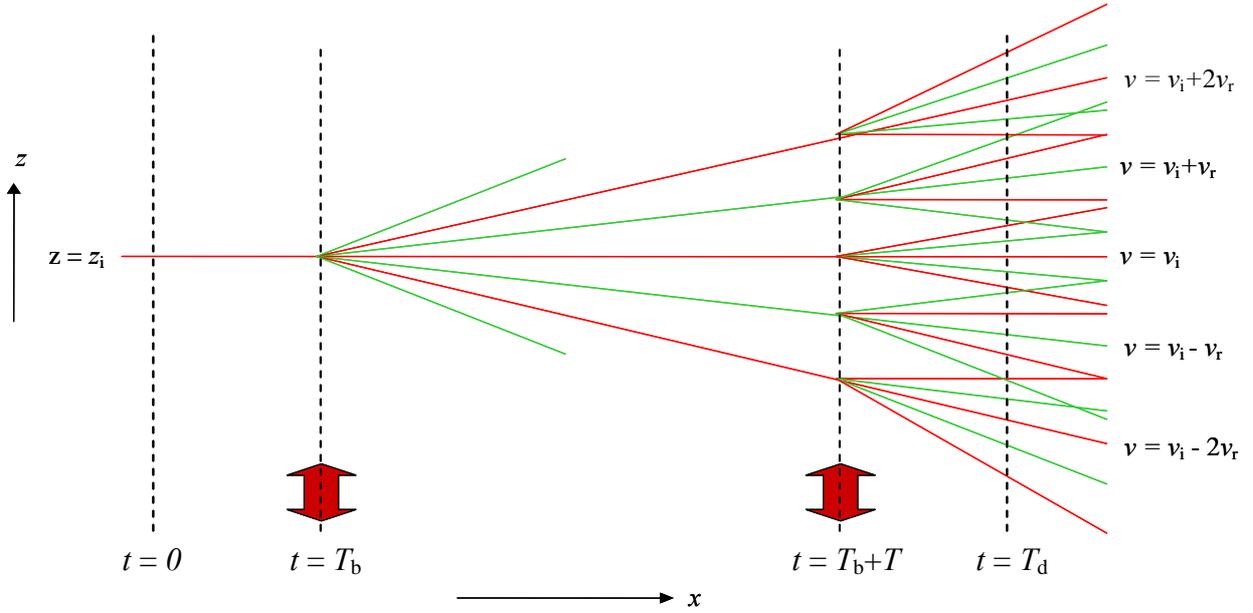

**Figure 1:** A two zone standing wave interferometer in the *x-z* plane. The atomic wave packet centres follow the red (ground state) and green (excited state) lines. Initially (*t* = 0) the atom is in the ground state at $z = z_i$ with a velocity component $v_i$ in the z direction. A first standing wave pulse is applied at $t = T_b$ and a second after a free evolution time $T$ (at $t = T_b+T$). All ground (excited) wave packets interfere at detection ($t = T_d$).

___________________________________________________________________________________

For the first interaction $T_b - \tau/2 \leq t \leq T_b + \tau/2$ we start with a transformation

$$\breve{\varphi}_n^a(t,z) = \varphi_n^a(t,z) \exp\left[\frac{iMv_r^2}{\hbar}\left(anT - \frac{n^2}{2}T_b\right)\right]$$
$$\breve{e}_n^a / \breve{g}_n^a(t) = e_n^a / g_n^a(t) \exp\left[-\frac{iMv_r^2}{\hbar}\left(anT - \frac{n^2}{2}T_b\right)\right].$$
(24)

We neglect the recoil kinetic energy and the effects related to the motion of the wave packets during the interaction time $\tau$ (Raman-Nath approximation). This amounts to neglecting phase terms that are of order $M(nv_r)^2\tau/\hbar$ i.e. $\approx \tau/T$ smaller than the main recoil shifts accumulated during free propagation, and variations of the field over distances of order $v_r\tau$ ($< 10^{-9}$ m for typical Cs fountain parameters). With this approximation it is easily seen (using (2) and transforming according to (24)) that during the interaction ($T_b - \tau/2 \leq t \leq T_b + \tau/2$) we have

$$\breve{\varphi}_n^a(t,z) \approx \breve{\varphi}_{n-1}^a(t,z) e^{ikz}.$$
(25)

Then, equating terms of identical $\breve{\varphi}_n^a(t,z)$, (23) reduces to an infinite set of coupled differential equations (equivalent to eqs. (3.116) of [8])

$$i\dot{\breve{e}}_n^a(t) = \frac{\Omega(t)}{4} e^{-i\Delta t}\left(e^{ik(z_i+v_it)}\breve{g}_{n-1}^a(t) + e^{-ik(z_i+v_it)}\breve{g}_{n+1}^a(t)\right)$$
$$i\dot{\breve{g}}_n^a(t) = \frac{\Omega(t)}{4} e^{+i\Delta t}\left(e^{ik(z_i+v_it)}\breve{e}_{n-1}^a(t) + e^{-ik(z_i+v_it)}\breve{e}_{n+1}^a(t)\right).$$
(26)

The coefficient of each $\breve{\varphi}_n^a(t,z)$ is coupled to two others corresponding to the wave packets whose central momenta differ by $\pm \hbar k$. The equations (26) model, of course, the multiple photon interactions taking place in the standing wave. Their solution provides the complete description of the system after the interaction. The only initial non-zero coefficient is $\breve{g}_0^0(t_0) = 1$ which provides the initial conditions for the solution ($t_0 = T_b - \tau/2$).

For the resonant case ($\Delta = 0$), a constant field in the x direction ($\Omega(t) \to \Omega$), and $z_i = v_i = 0$, (26) has an analytical solution in terms of Bessel functions [8]. With the above initial conditions

$$\breve{e}^0_{2m+1}(t-t_0) = i(-1)^{m+1} J_{2m+1}\left(\tfrac{1}{2}(t-t_0)\Omega\right)$$
$$\breve{g}^0_{2m}(t-t_0) = (-1)^m J_{2m}\left(\tfrac{1}{2}(t-t_0)\Omega\right) \tag{27}$$

where the $J_m(x)$ are Bessel functions of the first kind and m is an integer running from $-\infty$ to $\infty$. The probability to detect an atom in the excited state after the interaction is then simply given by

$$P_e(t_0+\tau) = \int_{-\infty}^{\infty} dz \left|\sum_{n=-\infty}^{+\infty} \breve{e}^0_n(t_0+\tau)\breve{\varphi}^0_n(t_0+\tau,z)\right|^2 = \sin^2\left(\tfrac{1}{2}\Omega\tau\right) \tag{28}$$

where we have used (27) and standard identities of the Bessel functions. So Rabi oscillations are recovered, in particular a $\pi/2$ pulse ($\Omega\tau = \pi/2$) does indeed correspond to a transition probability of 0.5. For the general case ($\Delta, z_i, v_i \neq 0$, $\Omega \to \Omega(t)$), (26) has to be solved numerically which, raises no particular difficulties (see section 4).

Between the two interactions the coupling field is zero so only the $\varphi^a_n(t,z)$ evolve, the coefficients $e^a_n(t)/g^a_n(t)$ remain unchanged (as can be easily seen from (23) when setting $\Omega = 0$).

The second interaction ($T_b+T-\tau/2 \leq t \leq T_b+T+\tau/2$) is treated in a similar way as the first one. We first transform, transferring phases accumulated during the free propagation (recoil energy) from the Gaussians $\varphi^a_n(t,z)$ to the coefficients $e^a_n(t)/g^a_n(t)$

$$\tilde{\varphi}^a_n(t,z) = \breve{\varphi}^a_n(t,z)\exp\left[\frac{iMv_r^2}{\hbar}\frac{n^2}{2}T\right]$$
$$\tilde{e}^a_n\big/\tilde{g}^a_n(t) = \breve{e}^a_n\big/\breve{g}^a_n(t)\exp\left[-\frac{iMv_r^2}{\hbar}\frac{n^2}{2}T\right] \tag{29}$$

The approximate relation (25) then becomes (under the same approximation and for $T_b+T-\tau/2 \leq t \leq T_b+T+\tau/2$)

$$\tilde{\varphi}^a_n(t,z) \approx \tilde{\varphi}^{a+1}_{n-1}(t,z)e^{ikz} \tag{30}$$

relating wave packets of different $v_i$ but same position at the second interaction. This leads to coupled equations analogous to (26)

$$i\dot{\tilde{e}}^a_n(t) = \frac{\Omega(t)}{4}e^{-i\Delta t}\left(e^{ik(z_i+v_it)}\tilde{g}^{a+1}_{n-1}(t) + e^{-ik(z_i+v_it)}\tilde{g}^{a-1}_{n+1}(t)\right)$$
$$i\dot{\tilde{g}}^a_n(t) = \frac{\Omega(t)}{4}e^{+i\Delta t}\left(e^{ik(z_i+v_it)}\tilde{e}^{a+1}_{n-1}(t) + e^{-ik(z_i+v_it)}\tilde{e}^{a-1}_{n+1}(t)\right) \tag{31}$$

which have to be solved once for each point at which one of the wave packets populated during the first interaction arrives (see figure 1) which provides the initial conditions for the solutions.

Finally, the solutions of (26) and (31) provide the coefficients $e^a_n(T_d)/g^a_n(T_d)$ at the time of detection and the probability of detecting either one of the internal states is the result of the interference between all wave packets in that internal state at detection. For example, for the excited state

$$P_e(T_d) = \int_{det.} dz \left|\sum_{n,a} e^a_n(T_d)\varphi^a_n(T_d,z)\right|^2 = \int_{det.} dz \left|\sum_{n,a} \Pi^a_n(T_d,z)\right|^2 \tag{32}$$

where the integral is taken over the spatial extension of the detection region and with the shorthand notation $\Pi^a_n(t,z) = e^a_n(t)\varphi^a_n(t,z)$.

The integral (32) provides the detection probability for an individual Gaussian wave packet (2) centred on initial position $z_i$ and velocity $v_i$. To obtain the actually observed detection probability the result of (32) has to be integrated over the statistical distributions ((12) and (14)) of central velocities and positions.

## 3. WEAK FIELD APPROXIMATION

In this section we perform an analytical calculation for the weak field case where interaction times and field intensities are such that all multiple photon interactions in the standing waves can be neglected i.e. in each interaction region the central momenta of the wave packets change by $\pm\hbar\mathbf{k}$ only. Then, looking at the detection probability for the excited internal state, only four states are populated and interfere at detection: $\Pi_1^0$, $\Pi_1^{-1}$, $\Pi_{-1}^1$, and $\Pi_{-1}^0$ (see figure 2). We calculate the interference and resulting detection probability arising from these four states.

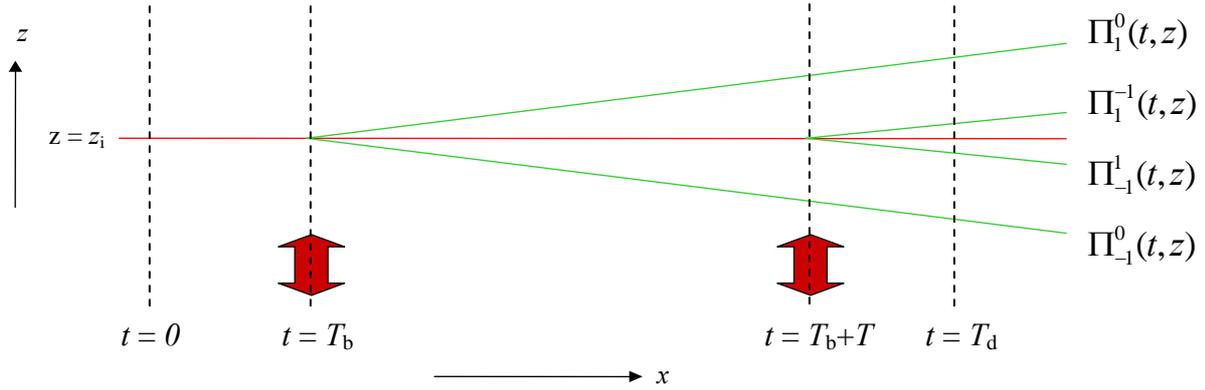

**Figure 2:** A two zone standing wave interferometer in the weak field approximation. Only four excited state wave packets interfere at detection.
_________________________________________________________________________________

We use similar assumptions as before, in particular we neglect recoil shifts accumulated during the interaction (Raman-Nath approximation) and treat the movement of the atoms in the x-dimension classically, with the same numerical consequences of these approximations as mentioned above. Furthermore, we assume that the detection region is much larger than the width of the wave packets at detection so the integral (32) can be extended from -∞ to ∞.

Although solutions can be found for the realistic cases of square or sinusoidal profiles of the field in the x direction ($V(x)$ in (21)) the solutions take their simplest form for a Gaussian profile, without changing the final result significantly (for calculations taking into account spatial and temporal dependences of the field see [24]). Using such a form for $V(x)$ leads to a time dependence of $\Omega(t)$ in (26) and (31) of the form

$$\Omega(t) = \Omega_0\, e^{-\frac{2(t-t_j)^2}{\tau^2}} \tag{33}$$

where $\Omega_0$ is the Rabi frequency at the maximum (cavity centre), $\tau$ is the interaction duration and $t_j$ is the central time of the interaction ($t_j = T_b$ for the first interaction and $t_j = T_b+T$ for the second one). For the first interaction the first order solution of (26) is given by the integral

$$\begin{aligned}
\breve{e}_{\pm 1}^0 &= \int_{-\infty}^{\infty} -i\frac{\Omega_0}{4} g_0^0 e^{\pm ikz_i} e^{-\frac{2(t-T_b)^2}{\tau^2}} e^{-i(\Delta \mp kv_i)t}\, dt \\
&= -i\sqrt{\frac{\pi}{2}}\frac{\Omega_0 \tau}{4} g_0^0 \exp\left[-\frac{(\Delta \mp kv_i)^2 \tau^2}{8}\right]\exp[i(-\Delta T_b \pm kz_i \pm kv_i T_b)]
\end{aligned} \tag{34}$$

and similarly for the second interaction the integral of (31) gives

$$\tilde{e}_{\pm 1}^{\mp 1} = -i\sqrt{\frac{\pi}{2}}\frac{\Omega_0 \tau}{4} g_0^0 \exp\left[-\frac{(\Delta \mp kv_i)^2 \tau^2}{8}\right]\exp[i(-\Delta(T_b + T) \pm kz_i \pm kv_i(T_b + T))] \tag{35}$$

where $g_0^0 = 1$ in both (34) and (35).

We then transform $\breve{e}_n^a / \tilde{e}_n^a$ back to $e_n^a$ using (24) and (29) and calculate the total detection probability from (32). The only interference terms in the sum of (32) that depend on $\Delta$ and can lead to a frequency shift are $\mathrm{Re}[(\Pi_{\pm 1}^0(z,T_d))^*(\Pi_{\pm 1}^{\mp 1}(z,T_d))]$ and $\mathrm{Re}[(\Pi_{\pm 1}^0(z,T_d))^*(\Pi_{\mp 1}^{\pm 1}(z,T_d))]$ which represent interference between wave packets moving in the same and opposite directions respectively. Substituting the $e_n^a$ from above, and $\varphi_n^a(z,T_d)$ from (2) and evaluating the z-integral we obtain

$$\mathrm{Re}[(\Pi_{\pm 1}^0(T_d))^*(\Pi_{\pm 1}^{\mp 1}(T_d))] \propto \exp\left[-\frac{(\Delta \mp kv_i)^2 \tau^2}{4}\right]\exp\left[-\frac{\hbar^2 k^2 T^2}{8M^2(\Delta z)^2}\right]\cos[\mp kv_i T + \Delta T - \delta T] \tag{36}$$

and

$$\mathrm{Re}[(\Pi_{\pm 1}^0(T_d))^*(\Pi_{\mp 1}^{\pm 1}(T_d))] \propto \exp\left[-\frac{(\Delta^2 + k^2 v_i^2)\tau^2}{4}\right]\exp\left[-2(\Delta z)^2 k^2 - \frac{\hbar^2 k^2}{2M^2(\Delta z)^2}\left(T_b + \frac{T}{2}\right)^2\right]$$
$$\times \cos[\pm 2kz_i \pm 2kv_i T_b \pm kv_i T + \Delta T + \delta T] \tag{37}$$

where $\delta = \hbar k^2/2M$ (the recoil shift) and the proportionality constant is the same in (36) and (37). So for a single Gaussian wave packet with initial central position $z_i$ and velocity $v_i$ in the rest frame of the cavity the detection probability of the excited state is proportional to the sum of (36) and (37). For $z_i = v_i = 0$ the above expressions are equivalent to equation (30) of [9], and to equations (23) and (27) of [15].

We now want to calculate the detection probability for a statistical ensemble of such wave packets corresponding to our density matrix (8). To do so we integrate (36) and (37) over the distribution of $v_i$ and $z_i$ given by (12) and (14). The $v_i$ integral of (36) yields

$$\left\langle \mathrm{Re}[(\Pi_{\pm 1}^0(T_d))^*(\Pi_{\pm 1}^{\mp 1}(T_d))]\right\rangle = \int_{-\infty}^{\infty} q(v_i)\mathrm{Re}[(\Pi_{\pm 1}^0(T_d))^*(\Pi_{\pm 1}^{\mp 1}(T_d))]$$
$$\propto \exp\left[-\frac{k^2 T^2 + a\Delta^2 \tau^2}{4a + k^2 \tau^2}\right] \tag{38}$$
$$\times \exp\left[-\frac{\hbar^2 k^2 T^2}{8M^2(\Delta z)^2}\right]\cos\left[-\frac{k^2 \Delta T \tau^2}{4a + k^2 \tau^2} + \Delta T - \delta T\right]$$

where $q(v_i)$ is the statistical distribution of central velocities corresponding to (12), the pointed brackets stand for the average over that distribution, and $a = M^2 / \left(2Mk_B\theta - \frac{\hbar^2}{2(\Delta z)^2}\right)$. Similarly the $v_i$ integral of (37) gives

$$\left\langle \mathrm{Re}[(\Pi_{\pm 1}^0(T_d))^*(\Pi_{\mp 1}^{\pm 1}(T_d))]\right\rangle \propto \exp\left[-\frac{\Delta^2 \tau^2}{4}\right]\exp\left[-\frac{k^2(2T_b + T)^2}{4a + k^2 \tau^2}\right]$$
$$\times \exp\left[-2(\Delta z)^2 k^2 - \frac{\hbar^2 k^2}{2M^2(\Delta z)^2}\left(T_b + \frac{T}{2}\right)^2\right]\cos[\pm 2kz_i + \Delta T + \delta T] \tag{39}$$

As noted earlier (section 1) we necessarily have $(\Delta z)^2 \geq \hbar^2/(4Mk_B\theta)$ so $a \geq M/(2k_B\theta)$. For Cs atoms at µK temperatures this implies $a \geq 7000$ (s/m)$^2$ while in typical experimental conditions $(k\tau)^2 \approx 1$ (s/m)$^2$, so we will expand (38) and (39) in $k^2\tau^2/4a$ and keep only leading terms. The two expressions simplify under that approximation to

$$\left\langle \text{Re}\left[\left(\Pi_{\pm 1}^0(T_d)\right)^*\left(\Pi_{\pm 1}^{\mp 1}(T_d)\right)\right]\right\rangle \propto \exp\left[-\frac{\Delta^2\tau^2}{4}\right]\exp\left[-\frac{k^2 k_B\theta}{2M}T^2\right]\cos[\Delta T - \delta T] \quad (40)$$

and

$$\left\langle \text{Re}\left[\left(\Pi_{\pm 1}^0(T_d)\right)^*\left(\Pi_{\mp 1}^{\pm 1}(T_d)\right)\right]\right\rangle \propto \exp\left[-\frac{\Delta^2\tau^2}{4}\right]\exp\left[-2(\Delta z)^2 k^2 - \frac{k^2 k_B\theta}{2M}(2T_b+T)^2\right] \\ \times \cos[\pm 2k z_i + \Delta T + \delta T] \quad (41)$$

Integrating (40) and (41) over the distribution of $z_i$ (14) we finally obtain

$$\left\langle\left\langle \text{Re}\left[\left(\Pi_{\pm 1}^0(T_d)\right)^*\left(\Pi_{\pm 1}^{\mp 1}(T_d)\right)\right]\right\rangle\right\rangle \propto \exp\left[-\frac{\Delta^2\tau^2}{4}\right]\exp\left[-\frac{k^2 k_B\theta}{2M}T^2\right]\cos[\Delta T - \delta T] \quad (42)$$

and

$$\left\langle\left\langle \text{Re}\left[\left(\Pi_{\pm 1}^0(T_d)\right)^*\left(\Pi_{\mp 1}^{\pm 1}(T_d)\right)\right]\right\rangle\right\rangle \propto \exp\left[-\frac{\Delta^2\tau^2}{4}\right]\exp\left[-2w^2 k^2 - \frac{k^2 k_B\theta}{2M}(2T_b+T)^2\right]\cos[\Delta T + \delta T] \quad (43)$$

where the double pointed brackets stand for the average over the statistical distributions of central velocities and initial positions ((12) and (14)). Note that (42) is unchanged from (40) as there is no $z_i$ dependence in (40).

Expressions (42) and (43) are the main results of the analytical part of our paper. They describe the interference between wave packets with equal or opposite recoil velocities (see figure 2). The total detection probability of the excited internal state is proportional to the sum of (42) and (43) which are affected by opposite recoil shifts, so the overall frequency shift will depend on the magnitudes of the contrast terms in (42) and (43) which involve the initial size of the atomic cloud $w$, its temperature $\theta$, the microwave wave vector $k$, and the characteristic times of the experiment $T_b$ and $T$. It is essential to note that neither of these equations depend on the unknown width of the individual wave packets ($\Delta z$) that enter the density matrix (8) so only the known (measured) quantities ($w, \theta$) play a physical role in the final observation.

Equation (42) describes the "usual" interference between co-propagating wave packets with a negative recoil shift $\delta T$ and the standard visibility function $\exp[-(v_r T)^2/(8 u^2)]$ for the interference of two Gaussian beams of waist $u = \hbar/(2\sqrt{k_B \theta M})$ separated by a distance $v_r T$. The second interference term (43) with positive recoil shift $\delta T$ has only been discovered recently [9, 10] and only plays a role when $2w^2 k^2$ is sufficiently small. This reflects the fact that in the integral over the detection region (32) the $\cos(2kz)$ term stemming from the opposite recoil directions of the two wave packets averages to zero unless the initial width of the cloud is of the same order or smaller than the wavelength of the electromagnetic field (so effectively the integral is carried out only over part of one wavelength). This is generally not the case for optical wavelength so (43) is insignificant in optical spectroscopy, but becomes significant for microwaves ($k \approx 135$ $m^{-1}$ inside a standard square cavity and $w \approx 1$ mm in a MOT). Also of interest is the second term in the second exponential of (43) which depends strongly on the time interval $T_b$ between the end of the cooling (focus of the atomic wave packets) and the first interaction, i.e. on the longitudinal coherence of the atomic source (distribution of $T_b$). For incoherent atoms, e.g. in a thermal beam (large distribution of $T_b$), this term leads to a zero average of (43). For a small distribution (a few 100 µs due to the adiabatic cooling phase) the term is not affected significantly (relative change of less than $10^{-3}$). Finally, we note that this second term can be driven to zero for $T_b = -T/2$ which corresponds to focusing the atomic cloud halfway between the two interactions (c.f. [15]) so (43) could become dominant with respect to (42).

In previous works [9, 10, 15] expressions (42) and (43) have been derived only for an individual Gaussian wave packet (equivalent to (36) and (37) with $z_i = v_i = 0$) which leads to results that depend on $\Delta z$ rather than $w$ and $\theta$ and may differ significantly from the ones obtained here using a statistical description based on the density operator (8). The expressions of [9, 15] are equivalent to (42) and (43) only for the pure state i.e. for $\hbar^2/(4Mk_B\theta) = (\Delta z)^2 = w^2$.

For microwave frequency standards the sum of (42) and (43) results in partial cancellation of the recoil shift depending on the particular experimental conditions. So the measured shift (see section 4 for numerical predictions) can be significantly smaller than the one expected from the order of magnitude (1) or from plane wave calculations (see section 4, table 1). Using the above analytical expressions it is easy to see that perfect cancellation occurs under the condition

$$\exp[-2w^2k^2 - 2k^2k_B\theta T_b(T_b+T)/M] = 1. \tag{44}$$

For standard caesium fountain parameters ($k \approx 135\ m^{-1}$, $\theta \approx 0.8\ \mu K$, $T_b \approx 0.15$ s, $T \approx 0.5$ s, $w \approx 1$ - 5 mm) the two terms in the exponential of (43) are of the same order and the exponential $\approx 0.81 - 0.34$. So we expect the observed recoil shift to depend on both terms and, in particular, on parameters ($w$, $\theta$, $T_b$, $T$) that are relatively easily varied in the experiment. But in all cases we expect it to be smaller than the plane wave prediction (equivalent to (43) = 0 and no visibility loss in (42)).

The analytical results obtained here only allow a qualitative understanding of the processes involved and are unlikely to provide the correct quantitative predictions for particular experimental conditions except for very special cases. This is due to the simplifying assumptions, in particular, the neglected multiple photon interactions at realistic field strengths that populate states at $n\hbar k$ with n>1, and to the extension of the detection region to $\pm\infty$ which both significantly affect the quantitative results. The next section presents a numerical simulation for realistic experimental conditions, the results of which are then compared to the analytical ones obtained here.

## 4. STRONG FIELD RESULTS

In this section we numerically calculate the frequency shift and fringe contrast of the observed Ramsey fringes in realistic experimental conditions. We use parameters of a "typical" Cs fountain with a rectangular microwave cavity, so $f(\mathbf{r})$ inside the cavity is described by (21) which leads to

$$\Omega(t) = \Omega_0 \cos(k_x v_x (t - t_j)) \tag{45}$$

in (26) and (31) with $t_j = T_b$ for the first interaction and $t_j = T_b+T$ for the second one. We substitute this into the sets of coupled equations (26) and (31) and solve them numerically, imposing a cut-off number of total photon exchanges (number of recoils $N_{rec}$) determined by the required accuracy of the solution and the microwave power (for example, we use $N_{rec} = 9$ for $\pi/2$ pulses). We then calculate the resulting detection probabilities via numerical integration of (32) over the detection region which is determined essentially by the size of the hole in the cavity.

To carry out the numerical solutions of (26) and (31) and the integral of (32) we need to fix the values of $\Delta z$, $v_i$, and $z_i$. To do so we make use of our weak field results (section 3) which showed that the chosen value of $\Delta z$ in the density operator (8) plays no physical role in the final observations. So we can choose any value provided we then integrate the result over the statistical distributions of $v_i$ (12) and $z_i$ (14). We choose $\Delta z = \hbar/(2\sqrt{k_B\theta M})$ so the distribution of $v_i$ becomes a delta function centred on $v_i = 0$. As a consequence we set $v_i = 0$ in our numerical solution and repeat the calculation for a set of values of $z_i$ drawn from the Gaussian distribution (14), with a cut-off ($\approx$ size of the hole in the cavity) which we impose in order to model the effect of the first cavity passage of the atoms. The final detection probabilities $\overline{P}_{e/g}(T_d)$ are then the averages of the ones obtained for the individual $z_i$.

Finally, the observables $O_{e/g}$ actually used for the determination of the shift are normalised to the total detection probability (similar to the routine operation of fountain clocks)

$$O_{e/g} = \frac{\overline{P}_{e/g}(T_d)}{\overline{P}_e(T_d) + \overline{P}_g(T_d)} \tag{46}$$

with the fringe contrast defined as $C = |O_e - O_g|$ at resonance ($\Delta = 0$).

The standard values of the parameters of the modelled fountain are as follows: total time of flight of the atoms, $T_d = 0.8$ s (total launching height $\approx 78$ cm); time from the end of cooling to the first interaction, $T_b = 0.15$ s (cavity $\approx 47$ cm above cooling region); time between the two interactions $T = 0.5$ s; horizontal cross-section of the detection region $\approx 1 \times 0.4$ cm. The atoms are cooled to a temperature of 0.8 µK with the measured initial size of the cloud $w = 1$ mm when operating in a MOT and $w = 5$ mm in a molasses. The cavity is rectangular of standard dimensions (22.86 x 10.16 mm cross-section) with its long edge parallel to the atom trajectory (aligned with the x-axis). So the standing wave inside the cavity is described by (21) where $V(x) = \cos(k_x x)$ with $k_x = 137.43$ m$^{-1}$ and $k = 135.04$ m$^{-1}$. We calculate the frequency shift and fringe contrast as a function of the initial width $w$ for a range of microwave powers, two launching heights (78 cm (T = 0.5 s) and 55 cm (T = 0.25 s)), and two temperatures (0.8 µK and 3.2 µK). The results are shown in figures 3 for the frequency shift and figure 4 for the contrast.

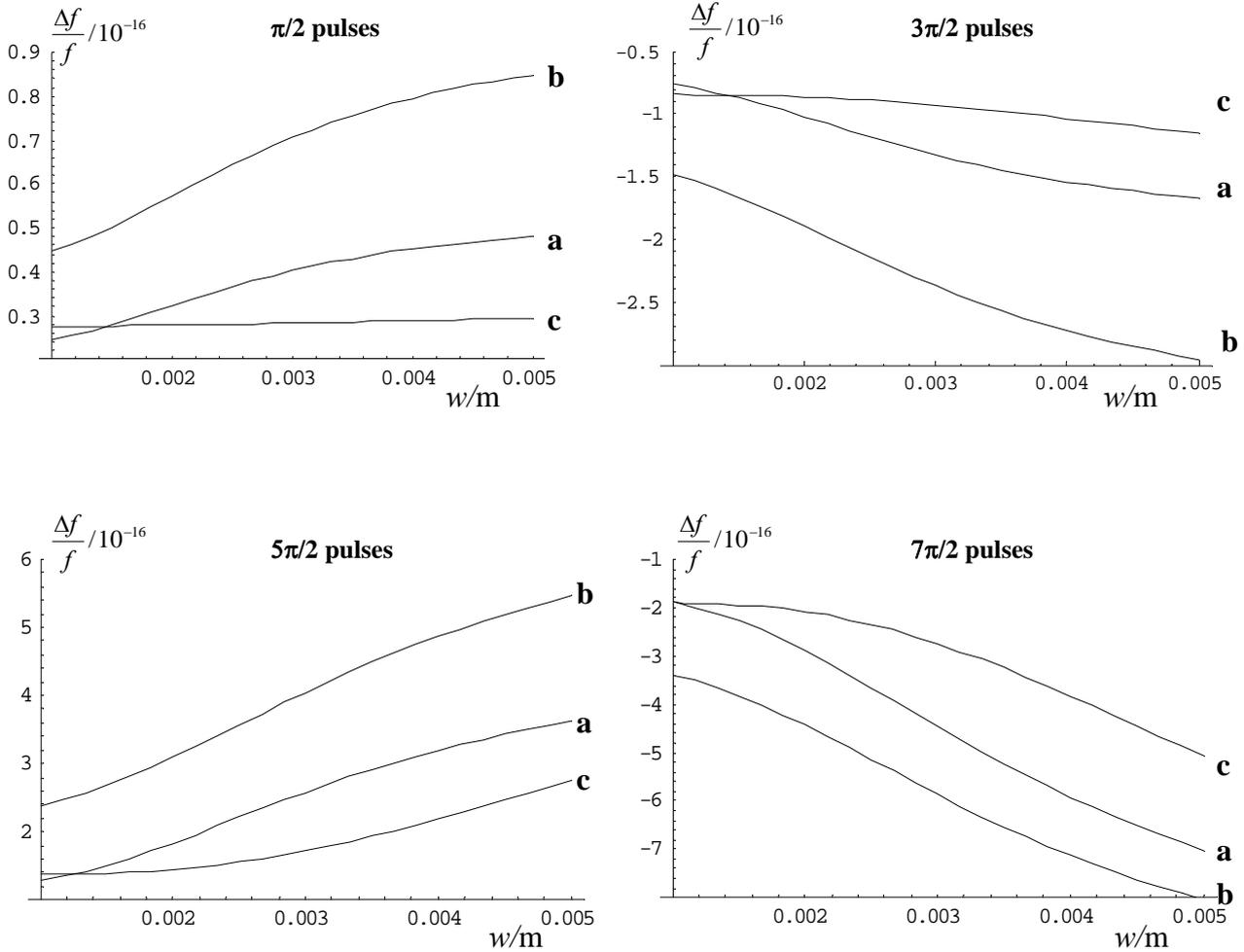

**Figures 3:** Numerical results for the recoil frequency shift as a function of the initial width of the atomic cloud ($w$). Typically $w = 1$ mm corresponds to a MOT while for atomic molasses $w$ can reach up to 5 mm. The four figures correspond to different microwave powers ($\pi/2$ to $7\pi/2$ pulses) with the three curves (a, b, c) representing different launching heights and temperatures. Curves "**a**" correspond to "standard" conditions i.e. launching height $\approx 78$ cm ($T_b = 0.15$ s, $T = 0.5$ s) and temperature $\theta \approx 0.8$ µK, "**b**" corresponds to lower height $\approx 55$ cm ($T_b = 0.21$ s, $T = 0.25$ s), and "**c**" to increased temperature $\theta \approx 3.2$ µK.

---

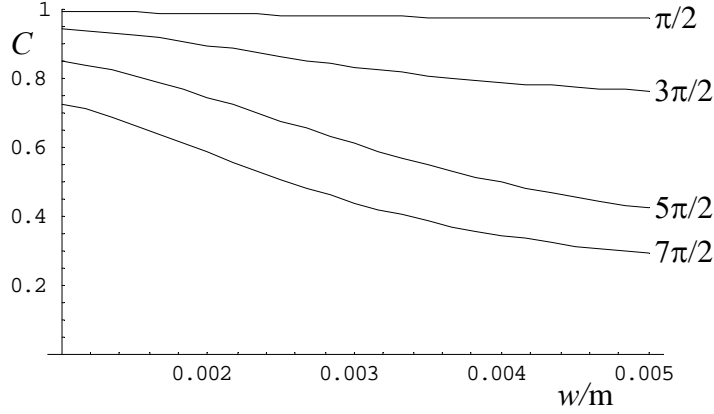

**Figure 4:** Numerical results for the fringe contrast ($C$) as a function of $w$ for different values of microwave power under "standard" conditions (launching height $\approx$ 78 cm, $\theta \approx 0.8$ μK).

___

One possible estimation of the accuracy of the numerical solutions of (26) and (31) is to check that the total detection probability when taking into account only the coefficients $\tilde{e}(T_d)/\tilde{g}(T_d)$ (plane wave approximation) sums to 1, i.e.

$$P_e^{PW}(T_d) + P_g^{PW}(T_d) = \left|\sum_{a,n}\tilde{e}_n^a(T_d)\right|^2 + \left|\sum_{a,n}\tilde{g}_n^a(T_d)\right|^2 = 1+\varepsilon \qquad (47)$$

where we generally obtain $\varepsilon \approx 10^{-7}$ which corresponds to an error in the calculated frequency shift of $\leq 10^{-17}$ in relative frequency. Table 1 gives the calculated relative frequency shift as a function of microwave power in the plane wave approximation ($P_e^{PW}(T_d)$ and $P_g^{PW}(T_d)$ defined as in (47) and $v_i = z_i = 0$). The contrast is always 100% in that case.

| Power ($\Omega\tau$) | Shift ($\Delta f /f$) |
|---|---|
| $\pi/2$ | 1.2 $10^{-16}$ |
| $3\pi/2$ | -3.6 $10^{-16}$ |
| $5\pi/2$ | 5.9 $10^{-16}$ |
| $7\pi/2$ | -8.3 $10^{-16}$ |

**Table 1:** "Plane wave" frequency shift as a function of microwave power.

___

All numerical integrals are carried out to at least 7 digit precision which leads to uncertainties that do not exceed 1 part in $10^{-17}$ in relative frequency in the calculated shifts. We have increased the number of values of $z_i$ drawn from the Gaussian distribution (14) until variations of that number by a factor 2 did not affect the results at the per cent level. The probably largest source of numerical error is due to the cut-off we impose on the drawn $z_i$ values in order to model the effect of the first cavity passage. Removing that cut-off completely, increases the resulting frequency shifts by up to 30 % for $w = 5$ mm but with no significant changes for $w = 1$ mm. Because of this we estimate our total numerical uncertainty to be about 10 % or less for large $w$ (molasses) and much smaller for $w = 1$mm (MOT).

As expected the plane wave shift (table 1) is greater than the order of magnitude estimated using (1) ($\approx$ 7.5 $10^{-17}$ with $k \approx 135$ m$^{-1}$ inside the cavity) because of the population of external states at $nv_r$ due to multiple photon processes in the standing waves. For the same reason the shift increases with microwave power as states with higher n get more populated. The change of sign can be easily understood qualitatively when one considers the states that are mainly populated for each value of $\Omega\tau$ and each internal state (c.f. the Bessel functions in (27)): for $\pi/2$ pulses the |g> state mainly populated is the initial state n=0 and the |e> state the one at n=±1. Hence the |e> states have (on average) higher kinetic energy than the |g> ones and the recoil shift is positive. On

the other hand, for 3π/2 pulses the |g> states mainly populated are those at n = ±2 but the |e> states are still those at n= ± 1 so the situation inverses.

The change of sign with increasing power is also observed for the recoil shifts calculated numerically using the complete Gaussian description of section 1 (c.f. figures 3), however the magnitudes are always significantly smaller than their plane wave counterparts. This is due to the partial cancellation of the shift in the sum of the interferences between wave packets with equal or opposite recoil directions as shown in equations (42) and (43) in the weak field approximation, with the condition for perfect cancellation in that approximation given in (44). The first term in the exponential of (44) explains qualitatively why the calculated shift increases with the initial width $w$ of the atomic cloud for all curves in figures 3. To test the influence of the second term we have run the strong field calculations for $T_b = 0$ (for π/2 pulses) whilst leaving all other parameters unchanged. As a result the frequency shift was decreased to 0.3 10$^{-17}$ (from 2.4 10$^{-17}$) for $w$ = 1 mm and to 3.8 10$^{-17}$ (from 4.8 10$^{-17}$) for $w$ = 5 mm, so the second term plays an important role, especially for small $w$, as expected.

However, the weak field results cannot explain, even qualitatively, the significant decrease of the shift at high temperature or its increase at lower launching height (curves b and c in figures 3). At increasing temperature one would expect the opposite effect from (44), as larger $\theta$ leads to a decrease of the exponential in (44) and therefore to a larger recoil shift. Similarly at lower launching height we have $T_b = 0.21$ s and $T = 0.25$ s (0.15 s and 0.5 s at standard height) so the second term in the exponential (44), and therefore the expected shift, remain almost unchanged. These differences clearly show the limits of the weak field approximation, even for a qualitative understanding of the numerical results for realistic experimental parameters. In both cases (higher temperature and lower height) the discrepancies are due to the fact that the detection region is limited to its realistic size (size of the hole in the cavity) in the numerical calculations (section 4) whereas it extends to ±∞ for the analytical results of section 3. In other words the analytical results reach their limit when the size of the atomic cloud at detection is comparable to, or larger, than the size of the detection region. This is confirmed by a decrease in the discrepancies when increasing the size of the detection region in the numerical calculations. For example, increasing the size of the detection region by a factor 2 leads to much more similar results for high and low launching heights (for π/2 pulses the remaining differences are ≤ 0.2 10$^{-17}$ (at $w$ = 1 mm) and ≤ 1 10$^{-17}$ (at $w$ = 5 mm) compared to 2.1 10$^{-17}$ and 3.7 10$^{-17}$ when using the standard detection region), and similarly at higher temperature (3.2 μK) the calculated shifts now approach and surpass those at standard temperature (0.8 μK).

The expected fringe contrast is shown in figure 4 which shows a loss of contrast with increasing microwave power and increasing initial width of the atomic cloud $w$. It should be noted that we have not "optimised" the contrast as is often the case in experiments i.e. we use $\overline{\Omega}\tau = N\pi/2$ for all calculations where $\overline{\Omega}$ is the average Rabi frequency along the classical trajectory that crosses the cavity in the centre (i.e. we use $\Omega_0 = N\pi^2/4\tau$ in (45)). As the atomic wave packets are distributed over a finite region of the cavity the power that optimises the contrast would be slightly bigger, especially at higher $N$ (higher microwave power) and for large $w$. As is well known, classical calculations also predict a loss of contrast as a function of the experimental parameters ($N$, $w$, $\theta$, $k$, $T$, $T_b$, etc.) and are confirmed by experiments [25]. In such calculations the atoms are treated as point particles following classical trajectories determined by their initial position and velocity distribution, and the fields only act on the internal degrees of freedom with the interaction times and field strengths for each individual atom determined by its trajectory. We have carried out such calculations for a given set of experimental parameters and found that they yield identical results (at the % level) to the ones we obtained using the methods described above. Therefore, unambiguous observation of the microwave recoil will require observation of the predicted frequency shifts (figures 3) which are not predicted by classical theory, additionally to the observation of the predicted contrast variations (figure 4).

## CONCLUSION

We have described the atomic cloud in cold atom interferometers in terms of a statistical ensemble of identical Gaussian wave packets and used that description to calculate the effect of the external degrees of freedom (in particular the recoil kinetic energy) of the atoms on the observations in microwave Ramsey spectroscopy.

In this paper we use such a Gaussian description for the wave packets in the two spatial dimensions that are perpendicular to the atomic trajectory but a classical description in the dimension parallel to the atomic trajectory. A more rigorous approach which does not require such a classical approximation can be found in [8]. For the purposes of the present paper the two approaches lead to similar results and predictions for the observable effects in "standard", present day microwave frequency standards (section 4). A discussion of the relation between the two approaches in the context of recoil effects in microwave frequency standards will be the subject of a forthcoming publication.

In section 3 analytical expressions for the observed fringe contrast and frequency shift were derived in the weak field approximation. The main results of that section are the analytical expressions (42) and (43) that describe the interference between wave packets with equal or opposite recoil velocities. The latter are responsible for a partial cancellation of the recoil frequency shift under particular experimental conditions (as detailed at the end of section 3) which are in general satisfied in microwave spectroscopy with cold atoms but not in the optical domain due to the much larger optical wave vectors. This makes observation of the microwave recoil shift even more difficult than already expected from its small value ($10^{-16}$ compared to $10^{-11}$ in the optical domain). The partial cancellation of the recoil shift was only discovered recently [9, 10] where it was described for a single Gaussian wave packet with equations that are only equivalent to (42) and (43) for the pure state. In this paper we provide a description of that effect based on a statistical ensemble of such wave packets which express the cancellation condition in terms of known quantities ($\theta$, $w$), giving significantly different results from the expressions in [9, 10, 15] which depend on the waist of one individual wave packet ($\Delta z$) rather than the measured temperature ($\theta$) and initial width ($w$) of the atomic cloud.

In section 4 we have numerically simulated realistic experimental situations (strong field, finite detection region, etc.), with the predicted frequency shifts and fringe contrast shown in figures 3 and 4. These show clearly that the expected recoil frequency shift is significantly smaller than expected from a simple order of magnitude (c.f. (1)) or from plane wave calculations (c.f. table 1). We have qualitatively explained those predictions using the analytical results of the previous section, but also shown that the approximations used for the analytical calculations reach their limit, even for a qualitative understanding, when the size of the atomic cloud at detection is of the same order or larger than the detection region.

At present, microwave frequency standards (Cs and Rb fountain clocks) reach accuracies of one part in $10^{15}$ or slightly below [2 - 5] and should therefore be insensitive to the recoil shifts calculated in section 4 which are still about an order of magnitude smaller in standard operation ($\pi/2$ pulses). However, new methods for controlling the major systematic effects [6] and the advent of space based clocks like the ACES mission (planned for 2008 [7]) should push the uncertainties to the low $10^{-16}$ regime or below which will make exact calculations and calibrations of the recoil shift indispensable. For that reason an early observation of recoil effects would be desirable, for example by exploring the variations of the recoil shift as a function of microwave power, launching height, temperature and initial width of the atomic cloud as predicted in figures 3. Note that the expected shift increases significantly with microwave power, so the most promising strategy for observation would be to look for the expected variations with launching height, temperature and initial width at maximum power. The results presented here are intended to provide some guidance for such experiments by showing the order of magnitude and expected behaviour of the involved effects, but they cannot provide precise numerical predictions for particular experiments as those depend critically on the involved experimental parameters (cavity shape, characteristic times, temperature etc.).

We finish by discussing some related issues which have been studied recently or could be addressed in future work. Perturbations due to other external fields (gravitational and inertial fields) also need to be included (see e.g. [15, 26]) and are the subject of present work. The description of the atomic cloud (section 2) can be generalised to any form of localised wave packets other than Gaussian (any form of $\left|\varphi_{p_i z_i}(t)\right\rangle$ localised on $p_i$ and $z_i$) and any kind of statistical ensembles of such wave packets (any functions $q(p_i)$ and $s(z_i)$) that enter the density matrix (8)) with the subsequent calculations (light interactions, detection etc.) carried out in a similar way. However, such descriptions reach their limit when the atoms cannot be described by an ensemble of identical wave packets, i.e. when the observables of the atomic cloud (e.g. momentum - position distributions and correlation functions) cannot be correctly described by a density operator of the form (8). Another limit of our description is reached when recoil effects during the interactions can no longer be neglected, in which case one has to take into account the deformation of the wave packets due to the interactions. Most importantly, the numerical results presented here have been obtained for the case of an atomic fountain clock operating with a square microwave cavity (c.f. (21)) as used, for example, in [4]. However, many other fountain clocks use a cylindrical cavity, in which case the essentially 1D problem studied here (only one possible recoil axis) becomes a 2D one (recoils possible in the whole plane perpendicular to the atomic velocity) which renders the resulting models and calculations significantly more complex.

**Acknowledgements:** We gratefully acknowledge fruitful discussions with Sébastien Bize, André Clairon, Chad Fertig, Kurt Gibble, Arnaud Landragin, Philippe Laurent, and Pierre Lemonde. We particularly thank Kurt Gibble and Chad Fertig for providing data on contrast measurements in their Rb-fountain, and CNES for financial support of one of us (PW).